\documentclass[%
 reprint,
superscriptaddress,
showpacs,
 amsmath,amssymb,
 aps,
]{revtex4-1}

\usepackage{graphicx}
\usepackage{color}
\usepackage{verbatim}
\usepackage{subfigure}
\DeclareMathOperator{\sech}{sech}
\DeclareMathOperator{\arcsinh}{arcsinh}

\begin{document}

\preprint{APS/123-QED}

\title{Continuous-variable entanglement of two bright coherent states that never interacted} 

\author{David Barral} 
\email{Corresponding author: david.barral@c2n.upsaclay.fr}
\author{Nadia Belabas}
\author{Lorenzo M. Procopio}
\affiliation{Centre de Nanosciences et de Nanotechnologies C2N, CNRS, Universit\'e Paris-Saclay, Route de Nozay, 91460 Marcoussis, France}
\author{Virginia D'Auria}
\author{S\'ebastien Tanzilli}
\affiliation{Universit\'e C\^ote d'Azur, Institut de Physique de Nice (INPHYNI), CNRS UMR 7010, Parc Valrose, 06108 Nice Cedex 2, France}
\author{Kamel Bencheikh}
\author{Juan Ariel Levenson}
\affiliation{Centre de Nanosciences et de Nanotechnologies C2N, CNRS, Universit\'e Paris-Saclay, Route de Nozay, 91460 Marcoussis, France}

\begin{abstract} We study continuous-variable entanglement of bright quantum states in a pair of evanescently coupled nonlinear $\chi^{(2)}$ waveguides operating in the regime of degenerate down-conversion. We consider the case where only the energy of the nonlinearly generated fields is exchanged between the waveguides while the pump fields stay independently guided in each original waveguide. We show that this device, when operated in the depletion regime, entangles the two non-interacting bright pump modes due to a nonlinear cascade effect. It is also  shown that two-colour quadripartite entanglement can be produced when certain system parameters are appropriately set. This device works in the traveling-wave configuration, such that the generated quantum light shows a broad spectrum. The proposed device  can be easily realized with current technology and therefore stands as a good candidate for a source of bipartite or multipartite entangled states for the emerging field of optical continuous-variable quantum information processing.
\end{abstract}

\date{November 8, 2017}
\maketitle 

\section {\it Introduction.}

In recent years there has been an increasing interest in quantum information processing (QIP) with continuous variables (CV) \cite{Braunstein2005, Andersen2010}. 
In contrast to optical QIP with discrete variables, where qubits are coded by using discrete photon observables, CV-based quantum information is encoded in the fluctuations of the field quadratures that can take a continuous spectrum of eigenvalues. Entanglement of such quadratures for Gaussian states constitutes the primary resource of CV-QIP protocols. Interestingly, the latter take advantage from deterministic resources and unconditional operations as well as highly efficient homodyne detection. These features have made optical CV a thriving area of research: teleportation \cite{Furusawa1998},  three-colour entanglement \cite{Coelho2009}, one-way quantum computation \cite{Miwa2009}, long-distance quantum key distribution \cite{Jouguet2013} and entanglement distillation and swapping \cite{Takahashi2010} are some examples of breakthrough demonstrations reached in this framework.

These achievements have been mostly accomplished with bulk optics in table-top experiments. However, integrated optics (IO) is one of the strongest candidates for transferring these systems to real-world light-based QIP technologies \cite{Tanzilli2012}. Both discrete and continuous-variable quantum states of light can be generated, processed, and measured in practical, low cost, interlinked, and reconfigurable optical chips. IO displays features out of reach from bulk-optics analogs such as miniaturization, sub-wavelength stability, generation and manipulation of quantum light by means of enhanced nonlinear effects and the thermo-, electro- or strain-optic properties of the substrates, and the ability to integrate detectors on chip  \cite{Alibart2016, Dietrich2016, Silverstone2016}. Nevertheless, IO-based CV is a relatively young area of research. The progressive advance of materials technology pushes this field through the development of highly nonlinear and low-loss materials such as, among others, {lithium niobate}, {potassium titanyl phosphate} and {silicon}. Over the last few years important on-chip demonstrations have been reported, such as continuous-wave single-mode squeezing up to $1.8$ dB in a traveling-wave configuration \cite{Kaiser2016} and up to $2.9$ dB in a cavity resonator architecture \cite{Dutt2015, Stefszky2017}, or the entanglement of a pair of squeezed states remotely produced in cavities via integrated configurable directional couplers \cite{Masada2015}. 


Beyond integrated $\chi^{(2)}$ single-waveguide performances, the on-chip integration of different functionalities is of primary importance. In this regard, the directional coupler with a built-in nonlinearity is an emblematic device that deserves special attention \cite{Perina2000}. Actually, the first demonstrations of two-photon NOON state generation in  nonlinear directional couplers have been recently shown \cite{Kruse2015, Setzpfandt2016}. The purpose of this work is to study the CV quantum properties of light propagated in this device in both spontaneous parametric down-conversion (SPDC) and optical amplification regimes. Our first prediction is the entanglement of the pump fields. This is particularly interesting since we consider independent pumps for each waveguide and no direct energy transfer between them, i.e. no evanescent coupling nor cavity feedback. We unveil the physical origin of such entanglement as a cascading nonlinear phase mediated, through pump depletion, by the interacting signal waves. We further demonstrate that entanglement between pumps can coexist with signals' entanglement leading to two-colour quadripartite entanglement of the four propagating modes for a specific set of parameters. Strikingly, there is no bulk-optics analog to this IO device or, in other words, a nonlinear beam splitter does not exist. Other  theoretical proposals dealing with the entanglement of bright waves in couplers have been reported over the last years \cite{Olsen2005, Olivier2006, Mallon2008, Migdley2010, Guo2010, Wang2013}. Let us stress, however, that from a fundamental point of view, in these studies a linear coupling between the pump fields is considered, whereas in our proposal these modes never interact directly. Moreover, from a practical point of view, these devices are based on doubly-resonant optical cavities, which are narrowband stationary-wave based devices with a limited bandwidth of entanglement. This is in contrast with our proposal which is based on traveling-waves and shows much broader bandwidths in the continuous-wave regime, only limited by the phase matching acceptance (up to 10 THz in perodically poled lithium niobate (PPLN) waveguides \cite{Alibart2016}). Therefore, our approach stands as a good option for practical CV-based optical QIP. Likewise, another strength of our proposal is its simplicity. As will be shown, CV entanglement is generated over a large range of propagation lengths by only controlling the input powers and phases. Finally, we stress that this device can be realized with current technology, notably on lithium niobate \cite{Alibart2016}.


\begin{figure}[t]
\centering
\includegraphics[width=0.47\textwidth]{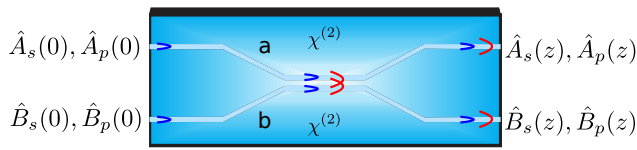}
\vspace {0cm}\,
\hspace{0cm}\caption{\label{F1}\small{(Colour online) Sketch of proposed nonlinear directional coupler made of two identical waveguides a and b with second-order susceptibilities $\chi^{(2)}$. Phase matching is fulfilled only in the coupling region. In red are the signal waves, evanescently coupled. In blue are the non-interacting pump waves.}}
\end {figure}

The article is organized as follows. In section II, we introduce the device under investigation and the propagation equations describing its operation. In section III we present the underlying cascade effect due to linear coupling of the signal modes that appears in the undepleted pump approximation and yields entanglement oscillations. In section IV we study the evolution of the covariance matrix, which describes the full system in the depleted pump regime, and show how bipartite and quadripartite entanglement arise. In section V we present the numerical results obtained for a specific example and discuss their importance. Finally, the main results of this work are summarized in section VI.

\section {The nonlinear directional coupler}

The nonlinear directional coupler, sketched in Figure \ref{F1}, is made of two identical $\chi^{(2)}$ waveguides in which  degenerate SPDC, or optical amplification when seeded, can take place. In each waveguide, a pump photon (p) at  frequency $\omega_{p}$ is down-converted into indistinguishable idler (i) and signal (s) photons with equal frequencies $\omega_{s,i}=\omega_{p}/2$ and the same polarization modes (fully degenerate process). The efficiency of this process is maximum when the mismatch of the propagation constants between the pump and signal photons caused by dispersion is negligible, so that $\Delta k \equiv k(\omega_{p})-2 k(\omega_{s})=0$. In the following we consider that this phase matching condition is fulfilled only in the coupling zone. The energy of the signal modes generated in each waveguide is exchanged between the waveguides through the evanescent waves, resulting into a linear coupling of the signal waves, whereas the interplay of the higher frequency pumps is negligible for the propagation lengths considered. {This approximation is made since the guided-modes fields are more confined into the guiding region as the modes wavelength decreases. As a result, the coupling constants for the pump modes are smaller than those corresponding to the signal modes \cite{Noda1981}}. The relevant operator which describes this system is the interaction momentum \cite{Linares2008, Barral2016}
\begin{equation} \label{Momentum}
\hat{M}=\hbar\, \{ g\,\hat{A}_{p} \hat{A}_{s}^{\dag\,2} + g\,\hat{B}_{p} \hat{B}_{s}^{\dag \,2} +C\,\hat{A}_{s} \hat{B}_{s}^{\dag} + h.c.\},
\end{equation}
where $\hat{A}$ and $\hat{B}$ are slowly varying amplitude annihilation operators of signal (s) and pump (p) photons corresponding to the upper (a) and lower (b) waveguides, respectively, $g$ is the nonlinear constant proportional to $\chi^{(2)}$, $C$ the linear coupling constant, $\hbar$ is Planck's  constant, and ${\it h.c.}$ stands for hermitian conjugate. From this momentum operator, the following Heisenberg equations are obtained
\begin{align} \label{QS1}
\frac{d \hat{A}_{s}}{d z}=& i C \hat{B}_{s} +2 i g \hat{A}_{p} \hat{A}_{s}^{\dag},   \\ \label{QS2}
\frac{d \hat{A}_{p}}{d z}=& i g \hat{A}_{s}^{2}, \\ \label{QS3}
\frac{d \hat{B}_{s}}{d z}=& i C \hat{A}_{s} +2 i g \hat{B}_{p} \hat{B}_{s}^{\dag},   \\
\frac{d \hat{B}_{p}}{d z}=& i g \hat{B}_{s}^{2},  \label{QS4}
\end{align}
where $z$ is the coordinate corresponding to the direction of propagation, and $C$ has been taken as real without loss of generality. Notably, the factor $2$ arising in both Equations (\ref{QS1}) and (\ref{QS3}) does not appear in the classically derived equations \cite{Linares2008}. 

\section {Undepleted-pumps approximation}

The production of CV entangled states in a nonlinear  $\chi^{(2)}$ directional coupler working in a traveling and continuous-wave configuration through SPDC has been thoroughly studied in ref.\,\cite{Herec2003}. In this case, the undepleted-pump approximation can in principle be safely assumed if strong coherent pumps are used, i.e., $ \vert \alpha_{p}\vert^{2}, \vert \beta_{p}\vert^{2}>>1$, where $\hat{A}_{p} \vert \alpha_{p} \rangle =  \alpha_{p} \vert \alpha_{p} \rangle$ and $\hat{B}_{p} \vert \beta_{p} \rangle =  \beta_{p} \vert \beta_{p} \rangle$. This previous work was focused on the analysis of the nonclassicality and entanglement of the quantum states related to the signal modes for different input parameters such as  pump powers and phases or coupling constants. It was then demonstrated that, if the linear coupling dominates over the nonlinear one ($C>2g$) for $\alpha_{p}=\beta_{p}$ (equal amplitudes and phases), a measure of entanglement, the logarithmic negativity $E_{\mathcal{N}}$ {(see section IV for a detailed discussion of this concept)}, shows an oscillatory evolution, periodically shifting between a maximum and zero values. This intriguing behaviour caught our attention. We solved analytically the system of Equations ($\ref{QS1}$) - ($\ref{QS4}$) in the undepleted-pump approximation, obtaining \cite{Perina2000}
\begin{align}
\hat{A}_{s}(z)=&\cos(\frac{\pi z}{2 L_{a b}}) \,\hat{A}_{s}(0) \nonumber \\
&+ i \frac{2 L_{a b}}{\pi}\sin(\frac{\pi z}{2 L_{a b}}) (2\eta \hat{A}_{s}^{\dag}(0)+ C \hat{B}_{s}(0) ), \label{US1} \\ 
\hat{B}_{s}(z)=&\cos(\frac{\pi z}{2 L_{a b}}) \,\hat{B}_{s}(0) \nonumber  \\
&+ i \frac{2 L_{a b}}{\pi}\sin(\frac{\pi z}{2 L_{a b}}) (2\eta \hat{B}_{s}^{\dag}(0)+ C \hat{A}_{s}(0) ), \label{US2}
\end{align}
where $L_{ab}=\pi / (2\sqrt{C^2 - 4\eta^2})$ is the beat length, i.e., the transfer distance of signal energy from waveguides a to b and from b to a, and $\eta\equiv g \vert \alpha_{p} \vert=g \vert \beta_{p} \vert$ acts as a nonlinear correction to the  linear coupling constant. To know if the absence of entanglement was related to energy conservation, we calculated the mean number of photons of the signal modes from the solutions given by Equations (\ref{US1})-(\ref{US2}), obtaining the following in each waveguide
\begin{equation}\label{Number}
N_{s}(z)=(\frac{4 \eta L_{ab}}{\pi})^2 \sin^{2} (\frac{\pi z}{2 L_{ab}} ).
\end{equation}
{Note that, in the case of no coupling, the usual $N_{s}(z)=\sinh^{2}(2\eta z)$ is found. Likewise, following ref.\,\cite{Herec2003} and using Equations (\ref{US1}) - (\ref{US2}) we calculated the evolution of the logarithmic negativity corresponding to the case under study, which is given by
\begin{equation}
E_{\mathcal{N}}(z)=-2\log_{2} (\sqrt{\sigma(z)-1/2}-\sqrt{\sigma(z)+1/2}\,),
\end{equation}
with $\sigma(z)=\sqrt{1+\left( ({C}/{\eta}) N_{s}(z) \right)^2}/2$}. Both maximum entanglement and number of signal photons are obtained at $z=(2n+1)L_{ab}$, with $n=0, 1, \dots$, whereas both signal photons and entanglement fully disappear at $z=2nL_{ab}$ \cite{Note0}. Since the intrinsic nature of the undepleted-pump approximation violates conservation of energy, this result indicates that the signal photons generated between $z=0$ and $z=L_{ab}$ are periodically transferred back to the pump in the interval $z=\{L_{ab}, \,2 L_{ab}\}$, but this behaviour is not accounted for in the undepleted-pump model \cite{Herec2003}. 

After this connection between the entanglement evolution and the conservation of energy, the next step is to identify the origin of this effect. Since perfect phase matching is considered in the system, we wonder if a dynamical phase is  showing up during the propagation. To tackle this question, we calculate the classical phase accumulated by the signal modes. Optical parametric amplification, i.e., seeding the signal modes, is considered since SPDC is not suitable to this calculation. Taking identical input seeds and calculating their propagation through Equations (\ref{US1})-(\ref{US2}), the following phase mismatch is obtained
\begin{equation} \label{PhiClass}
\Delta\Phi (C, \eta, z) = -2 \arctan\{\sqrt{\frac{C+2\eta}{C-2\eta} } \tan    (\frac{\pi z}{2 L_{ab}})       \},
\end{equation}
where $\Delta\Phi\equiv\Phi_{p}-2\Phi_{s}=-2\Phi_{s}$ with $\Phi_{s}$ being the classical phase related to each signal mode and $\Phi_{p}$ being a constant phase corresponding to the pump taken as a reference ($\Phi_{p}=0$). This nonlinear phase mismatch drives two cascaded nonlinear optical processes, down-conversion followed by up-conversion, mediated by the linear coupling of the signal modes. This evolving phase mismatch periodically switches the system from an efficient pump-to-signal conversion to an efficient signal-to-pump conversion. At the beat lengths, the cascaded phases are $\Delta\Phi (C, \eta, (2n+1)L_{ab})= \pi$ or $\Delta\Phi (C, \eta, 2n L_{ab})=2 \pi$. Applying these phases in Equation (\ref{Momentum}) is equivalent to switching the nonlinear coupling constant from $g$ to $-g$ at the odd multiples of $L_{ab}$ and keeping its sign positive at the even multiples. Therefore this phase mismatch $\Delta\Phi$ is at the origin of the transfer of photons from pumps to signals and from signals to pumps. Note that in the case of uncoupled waveguides, that is $C=0$, only down-conversion is produced in each waveguide and the above phase disappears due to the phase matching. 
 
In light of these results a question arises, if down-conversion within the nonlinear  coupler produces entanglement of the signal modes, does up-conversion entangle the pump modes which do not directly interact in the system? To answer this question we need to include the quantum character of the pump modes. This is carried out in the following section by means of the linearisation method in the depleted pump regime.

\section {Depleted-pumps regime}

Unfortunately, there is no known exact analytical solution to Equations (\ref{QS1}) - (\ref{QS4}) when the quantum nature of the pump fields is taken into account. We thus implement the linearisation of the equations by means of quantum-fluctuation operators with zero mean value and the same variances as the input operators $\hat{A}_{j}, \hat{B}_{j}$, with $j=s, p$, {which is a standard method in quantum optics} \cite{Reynaud1992}. 
Under this approximation, we firstly need to solve the propagation of the classical fields $\alpha_{s} (\alpha_{p})$ and $\beta_{s} (\beta_{p})$, in order to obtain the evolution of the quantum fluctuations. Let us define the following dimensionless amplitudes and phases related to the classical fields for each waveguide \cite{Li1994}
\begin{align} \label{DimVar}
u_{s}(z)&=\frac{\vert \alpha_{s}(z)\vert}{\sqrt{P}},  &  &v_{s}(z)=\frac{\vert \beta_{s}(z)\vert}{\sqrt{P}}, \\
\theta_{s}(z)&=\arg\{\alpha_{s}(z)\}, &  &\phi_{s}(z)=\arg\{\beta_{s}(z)\}, \\
u_{p}(z)&=\sqrt{\frac{2}{P}}\,\vert \alpha_{p}(z)\vert,  &  &v_{p}(z)= \sqrt{\frac{2}{P}}\,\vert \beta_{p}(z)\vert,\\
\theta_{p}(z)&=\arg\{\alpha_{p}(z)\},  &  &\phi_{p}(z)=\arg\{\beta_{p}(z)\},
\end{align}
and the normalized propagation coordinate $\zeta=\sqrt{2 P} g z$, with $P$ a constant of propagation related to the conservation of energy and the power of the whole system, which reads
\begin{equation}
P=\vert \alpha_{s}(z) \vert^{2} + \vert \beta_{s}(z) \vert^{2}  + 2\vert \alpha_{p}(z) \vert^{2} + 2\vert \beta_{p}(z) \vert^{2}, \quad  \forall z \geq 0.
\end{equation}
Introducing these variables into the classical version of Equations (\ref{QS1}) - (\ref{QS4}), we obtain  \cite{Li1994}
\begin{align}\label{us}
\frac{d {u}_{s}}{d \zeta}=&  -\kappa \, {v}_{s} \sin(\phi_{s}-\theta_{s}) - u_{s} u_{p} \sin(\Delta\theta),  \\
\frac{d {\theta}_{s}}{d \zeta}=& \kappa \frac{{v}_{s}}{u_{s}} \cos(\phi_{s}-\theta_{s}) + u_{p} \cos(\Delta\theta),\\
\frac{d {u}_{p}}{d \zeta}=& u_{s}^2 \sin(\Delta\theta), \\
\frac{d {\theta}_{p}}{d \zeta}=& \frac{u_{s}^2}{u_{p}} \cos(\Delta\theta), \label{tp}
\end{align}
with $\Delta\theta \equiv \theta_{p}-2\theta_{s}$ and $\kappa=C/(\sqrt{2P} g)$. The other four equations can be obtained by exchanging $u\leftrightarrow v$ and $\theta \leftrightarrow \phi$. $\kappa$ is the governing parameter of the system. It acts as an effective coupling which relates the linear and nonlinear couplings as well as the total input power. It is important to notice that we restrict the calculations to values of $\kappa>1$, since for $\kappa \leq 1$ the fluctuations start to grow exponentially and the linearisation approximation is no longer valid \cite{Mallon2008}. This situation corresponds to the experimental regime available with the current technology. Moreover, when considering identical input energy at each waveguide, the following initial conditions are chosen
\begin{gather}
u_{s}(0)=v_{s}(0)= \sech(-\delta_{0})/\sqrt{2},\\
u_{p}(0)=v_{p}(0)=-\tanh(-\delta_{0})/\sqrt{2}, 
\end{gather}
such that $\sum_{j=s,p} (u_{j}^2 + v_{j}^2)=1$ at any normalized plane $\zeta$. The initial conditions are related to the measurable signal and pump input powers, $P_{s}$ and $P_{p}$, through the parameter $\delta_{0}=\arcsinh(\sqrt{P_{p}/P_{s}})$. Note that in the case of different input powers in each waveguide a different set of initial conditions should be used. Furthermore, the behaviour of the light propagating in the device depends on the input phases $(\theta_{s}(0), \theta_{p}(0), \phi_{s}(0), \phi_{p}(0))$. However, we found that if all initial phases are set equal, the pump-fields entanglement is maximised. Therefore, $\theta_{s}(0)$=$\theta_{p}(0)$=$\phi_{s}(0)$=$\phi_{p}(0)$ is taken along the paper.

{The solutions of this classical system of equations are then fed into first order equations in the quantum fluctuations keeping only the linear terms}. Since we are interested in CV-entanglement, it is more convenient to deal with quadratures of the field $\hat{X}^{(A,B)}_{(s,p)}$, $\hat{Y}^{(A,B)}_{(s,p)}$, where $\hat{X}^{O}=(\hat{O}+\hat{O}^{\dag})/\sqrt{2}$ and $\hat{Y}^{O}=i (\hat{O}^{\dag}-\hat{O})/\sqrt{2}$ are the orthogonal quadratures corresponding to an optical mode ${O}$. In terms of dimensionless variables, the propagation of the quantum field quadratures are given by  \cite{Li1994}
\begin{align} \label{Quad1}
\frac{d \hat{X}_{s}^{A}}{d \zeta}= &- u_{p} \sin(\theta_{p}) \hat{X}_{s}^{A} + u_{p} \cos(\theta_{p}) \hat{Y}_{s}^{A}  - \kappa \hat{Y}_{s}^{B} \nonumber \\ 
&+ \sqrt{2} u_{s} \sin (\theta_{s}) \hat{X}_{p}^{A} - \sqrt{2} u_{s} \cos (\theta_{s}) \hat{Y}_{p}^{A},     \\  \label{Quad2}
\frac{d \hat{Y}_{s}^{A}}{d \zeta}=  &\, u_{p} \cos(\theta_{p}) \hat{X}_{s}^{A} + u_{p} \sin(\theta_{p}) \hat{Y}_{s}^{A}  + \kappa \hat{X}_{s}^{B}  \nonumber \\ 
&+ \sqrt{2} u_{s} \cos (\theta_{s}) \hat{X}_{p}^{A} + \sqrt{2} u_{s} \sin (\theta_{s}) \hat{Y}_{p}^{A},       \\   \label{Quad3}
\frac{d \hat{X}_{p}^{A}}{d \zeta}=    &-\sqrt{2} u_{s} \sin(\theta_{s}) \hat{X}_{s}^{A} - \sqrt{2} u_{s} \cos(\theta_{s}) \hat{Y}_{s}^{A},    \\
\frac{d \hat{Y}_{p}^{A}}{d \zeta}=  &\, \sqrt{2} u_{s} \cos(\theta_{s}) \hat{X}_{s}^{A} - \sqrt{2} u_{s} \sin(\theta_{s}) \hat{Y}_{s}^{A}, \label{Quad4}
\end{align}
and the other four equations are obtained by exchanging again $u\leftrightarrow v$, $\theta \leftrightarrow \phi$ and $A \leftrightarrow B$. This system of equations can be rewritten in compact form as $d \hat{\xi}/ d \zeta = \mathbf{\Delta}(\zeta)\, \hat{\xi}$, where $\mathbf{\Delta}(\zeta)$ is a $8\times8$ matrix of coefficients of Equations (\ref{Quad1}) - (\ref{Quad4}), and $\hat{\xi}=(\hat{X}_{s}^{A},\hat{Y}_{s}^{A},\hat{X}_{p}^{A},\hat{Y}_{p}^{A},\hat{X}_{s}^{B},\hat{Y}_{s}^{B},\hat{X}_{p}^{B},\hat{Y}_{p}^{B})^T$. The formal solution of this equation is given by
\begin{equation}
\hat{\xi}(\zeta)=\mathbf{S}(\zeta)\, \hat{\xi}(0),
\end{equation}
with $\mathbf{S}(\zeta)=\exp\{\int_{0}^{\zeta}\mathbf{\Delta}(\zeta')\, d \zeta' \}$. This is a linear canonical transformation between the input and output quadratures of the four fields which contains the full evolution of our quantum system. Remarkably, Equations (\ref{Quad3}) and (\ref{Quad4}) provide information about the propagation of the quantum fluctuations of the pumps in each waveguide which, as we will see, are entangled when the system is appropriately set.

Now, a suitable and experimentally accessible observable of the system has to be selected. Since the input states we deal with are Gaussian -actually coherent states- and since we are interested in CV entanglement properties of the four modes interacting within the system, we choose the second-order moments of the quadrature operators, properly arranged in the covariance matrix $\mathbf{V}$ with elements defined as \cite{Adesso2014, Buono2010}
\begin{equation}
V({\xi}_{j}, {\xi}_{k})=\frac{1}{2}(<\Delta\hat{\xi}_{j} \Delta\hat{\xi}_{k}> + <\Delta\hat{\xi}_{k} \Delta\hat{\xi}_{j}>),
\end{equation}
where $\Delta \hat{\xi}\equiv\hat{\xi}-\langle\hat{\xi} \rangle$. This is a real symmetric matrix that contains all the useful information about the quantum states propagating in the device. The covariance matrix corresponding to the input four-mode state, where the pumps are coherent states and the signals vacuum (spontaneous down-conversion) or coherent (optical amplification), is proportional to the identity matrix $\mathbf{V}(0)=(1/2)\, \mathbf{I}_{8}$, where 1/2 corresponds to the shot noise in our notation. Likewise, the covariance matrix at any normalized propagation plane $\zeta$ is given by
\begin{equation} \label{cov}
\mathbf{V}(\zeta)=\mathbf{S}(\zeta) \mathbf{V}(0) \mathbf{S}^{T}(\zeta).
\end{equation}
The positivity of $\mathbf{V}$ indicates that it is a {\it bona fide} covariance matrix. This feature is mathematically given by the condition $\mathbf{V}(\zeta)+i\mathbf{\Omega}/2 \geq 0$, equivalent to the Heisenberg uncertainty principle, where $\mathbf{\Omega}$ is the symplectic form given by $\mathbf{\Omega} = \oplus_{k=1}^4 \mathbf{\sigma}$ with $\mathbf{\sigma}\equiv$ adiag$[1,-1]$.
We have checked that this condition is satisfied in all the cases studied in this paper.

The covariance matrix can be efficiently measured by means of homodyne detection \cite{Buono2010, Dauria2009}. In these works, the covariance matrices of bipartite systems are considered. Once $\mathbf{V}$ is known, the amount of CV entanglement in bipartite systems is easily quantified through the logarithmic negativity $E_{\mathcal{N}}$ \cite{Vidal2002}. This entanglement witness is based on the {Peres-Horodecki-Simon} criterion, which establishes that a quantum state is entangled if the partially transposed density matrix is non-positive.
The logarithmic negativity is obtained from the symplectic spectrum $\{\nu_{k}\}_{k=1}^4$ of the partial transpose of the covariance matrix with respect to a subsystem $j$, $\mathbf{V}^{T_{j}}$, computed as the standard eigenvalues spectrum of the matrix $\vert i \mathbf{\Omega} \mathbf{V}^{T_{j}} \vert$. $E_{\mathcal{N}}$  is then given by \cite{Vidal2002}
\begin{equation}\label{EN}
E_{\mathcal{N}}=\sum_{k=1}^{4} F(\nu_{k}) \,\,\text{with}\,\, F(\nu)= 
\begin{cases}
    0       & \,\, \text{for}\,\, \nu\geq1/2 \\
    -\log_{2}(2\nu) & \,\, \text{for}\,\, \nu<1/2\\
  \end{cases}
\end{equation}
such that any value $E_{\mathcal{N}}>0$ indicates entanglement. Futhermore, in addition to entanglement quantification, this function also presents other appealing properties such as additivity. It also represents an upper bound on the efficiency of distillation, measuring to what extent a quantum state is useful for a certain QIP protocol or the number of resources needed to generate it \cite{Vidal2002}.

In the regime under study the system is composed of four quantum fields. Then, it would be also interesting to investigate if there is any collection of parameters for which multipartite entanglement is also displayed. The logarithmic negativity is limited to the measurement of bipartite entanglement, where each party is composed by one or more modes, but van Loock and Furusawa introduced a set of conditions to be simultaneously fulfilled for detecting multipartite full inseparability in CV systems \cite{vanLoock2003}. This criterion leads to genuine multipartite entanglement when pure states are considered. In the case of mixed states a more general criterion has been devised \cite{Shalm2013}. Since we deal with pure states, we use the criterion established in ref.\,\cite{vanLoock2003}. Transposed to our four-mode quantum state, genuine quadripartite entanglement is present if the following inequalities are simultaneously violated\begin{align} \label{VLF}
\hspace{-0.2cm} \langle [\Delta(\hat{X}_{s}^{A} - \hat{X}_{p}^{A})]^2 \rangle +  \langle [\Delta(\hat{Y}_{s}^{A} + \hat{Y}_{p}^{A}+r_{3} \hat{Y}_{s}^{B} + r_{4} \hat{Y}_{p}^{B})]^2 \rangle \geq 2,  \nonumber \\
\hspace{-0.2cm} \langle [\Delta(\hat{X}_{p}^{A} - \hat{X}_{s}^{B})]^2 \rangle +  \langle [\Delta(r_{1} \hat{Y}_{s}^{A} + \hat{Y}_{p}^{A}+\hat{Y}_{s}^{B} + r_{4} \hat{Y}_{p}^{B})]^2 \rangle \geq 2,  \nonumber \\
\hspace{-0.2cm} \langle [\Delta(\hat{X}_{s}^{B} - \hat{X}_{p}^{B})]^2 \rangle +  \langle [\Delta(r_{1} \hat{Y}_{s}^{A} + r_{2} \hat{Y}_{p}^{A}+ \hat{Y}_{s}^{B} + \hat{Y}_{p}^{B})]^2 \rangle \geq 2,
\end{align}
where $r_{j}$ (j=1,\dots,4) are arbitrary real parameters used for optimisation. These equations can be easily rewritten in terms of the elements of the covariance matrix, being therefore calculated from Equation (\ref{cov}) \cite{Migdley2010}. 

Consequently, from the covariance matrix $\mathbf{V}$ related to the multimode quantum system under study, we can measure the bipartite entanglement and assess if quadripartite entanglement is present with the tools introduced above.

\section{Numerical results and discussion}

We now present the numerical results obtained for a specific example with the theory developed in section IV. We display a detailed study of the CV bipartite entanglement in two regimes, negligible depletion (or undepletion) and high depletion, and quantify it by means of the logarithmic negativity given by Equation (\ref{EN}). Then we assess the potential of producing  quadripartite entanglement with this device by means of the van Loock and Furusawa inequalities given by Equations (\ref{VLF}), and finally we present a discussion about the feasibility {and possible applications} of the investigated device with current technology.

As device substrate we choose {lithium niobate} due to its appealing properties such as low propagation losses, large conversion efficiencies and broad bandwidth when operated in continuous wave \cite{Alibart2016}. The conversion efficiency of the nonlinear process can be maximised mainly in two ways. First, the input pumps are set as strong coherent fields generating type-0 down-conversion, which couples pump and signal modes polarized along the extraordinary (e) axis of the crystal through the higher component of the second order nonlinear tensor ($d_{33}\approx 34$ pm/V). Second, a periodically poled grating is engineered along the substrate (PPLN) in order to obtain quasi-phase matching of the propagation constants. The periodicity of the grating $\Lambda$, is such that the  mismatch $\Delta k$ is compensated by the wavevector associated with the grating  $\Delta k \approx 2\pi / \Lambda$. Perfect or partial quasi-phase matching has been considered at the quantum level in different theoretical articles, where it was demonstrated that the linearisation procedure also applies for PPLN \cite{Bencheikh1995, Noirie1997}. Typical PPLN chip lengths are on the order of a few centimeters.

We have solved numerically the classical mean values Equations (\ref{us}) - (\ref{tp}), the quantum quadratures Equations (\ref{Quad1}) - (\ref{Quad4}) and the propagation of the covariance matrix Equation (\ref{cov}) for the following waveguide parameters, $C=8.\,10^{-2}$ \,mm$^{-1}$ and $g=25.\,10^{-4}$ \,mm$^{-1}$ mW$^{-1/2}$, which will be used in the remainder of the article. These are standard values in PPLN waveguides \cite{Alibart2016, Kaiser2016}. Particularly, the value of linear coupling is chosen sufficiently low in order to obtain large nonlinear effects in the considered PPLN lengths, but similar results are obtained with other realistic values of $C$ and $g$.

\subsection{Bipartite entanglement}

In Figures \ref{F2} and \ref{F3} we show, respectively, for the negligible and high depleted-pump fields regimes, the dimensionless classical powers for each mode in each waveguide along the propagation (Figures \ref{F2}a-\ref{F3}a), relevant elements of the correlation matrix, $V(X_{(s, p)}^{A},X_{(s,p)}^{B})$ and $V(Y_{(s, p)}^{A},Y_{(s, p)}^{B})$ (see appendix), related to the signal (s) and pump (p) subsystems (Figures \ref{F2}b-\ref{F3}b and \ref{F2}c-\ref{F3}c, respectively), and  the logarithmic negativity $E_{\mathcal{N}}$ corresponding to the signals and pumps which quantifies bipartite entanglement (Figures \ref{F2}d-\ref{F3}d). The following effective coupling is chosen to be $\kappa=1.13$, in such a way that $\zeta=1$ stands for $z\approx 14$ mm. Likewise, the ratio between the signal and pump powers at each waveguide is taken as $P_{s}/ P_{p}=10^{-20}$ for the negligible depletion case \cite{Note1} and as $P_{s} / P_{p}=1$ for the high depletion one, fixing $\delta_{0}$ for each case.

\subsubsection{Negligible depletion}

\begin{figure}[p]
  \centering
    \subfigure{\includegraphics[width=0.42\textwidth]{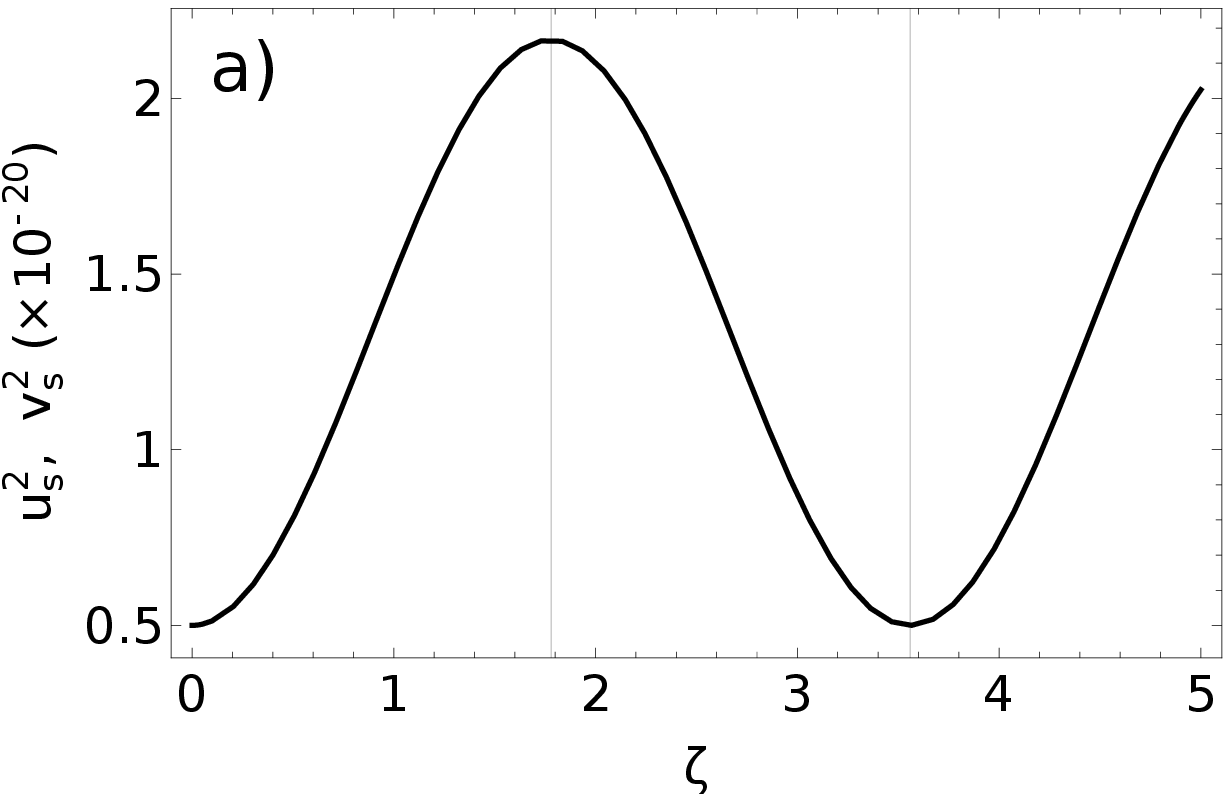}}
    \vspace {-0.2cm}
    \subfigure{\includegraphics[width=0.43\textwidth]{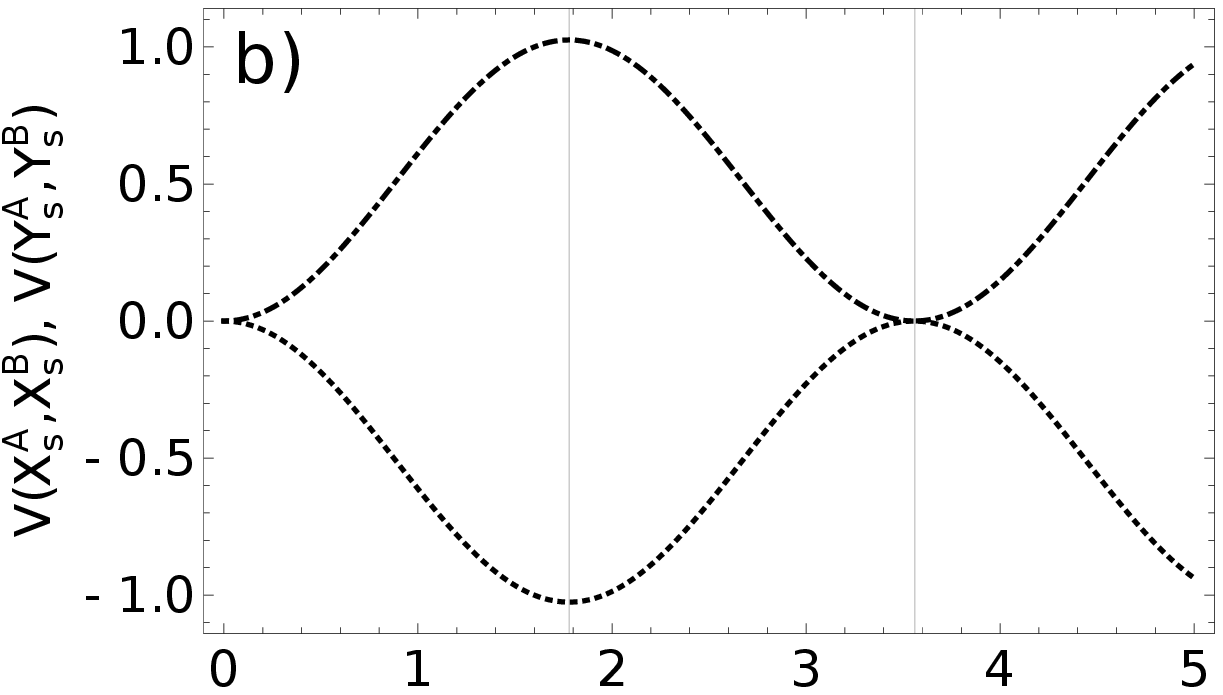}}
    \vspace {-0.2cm}
     \subfigure{\includegraphics[width=0.44\textwidth]{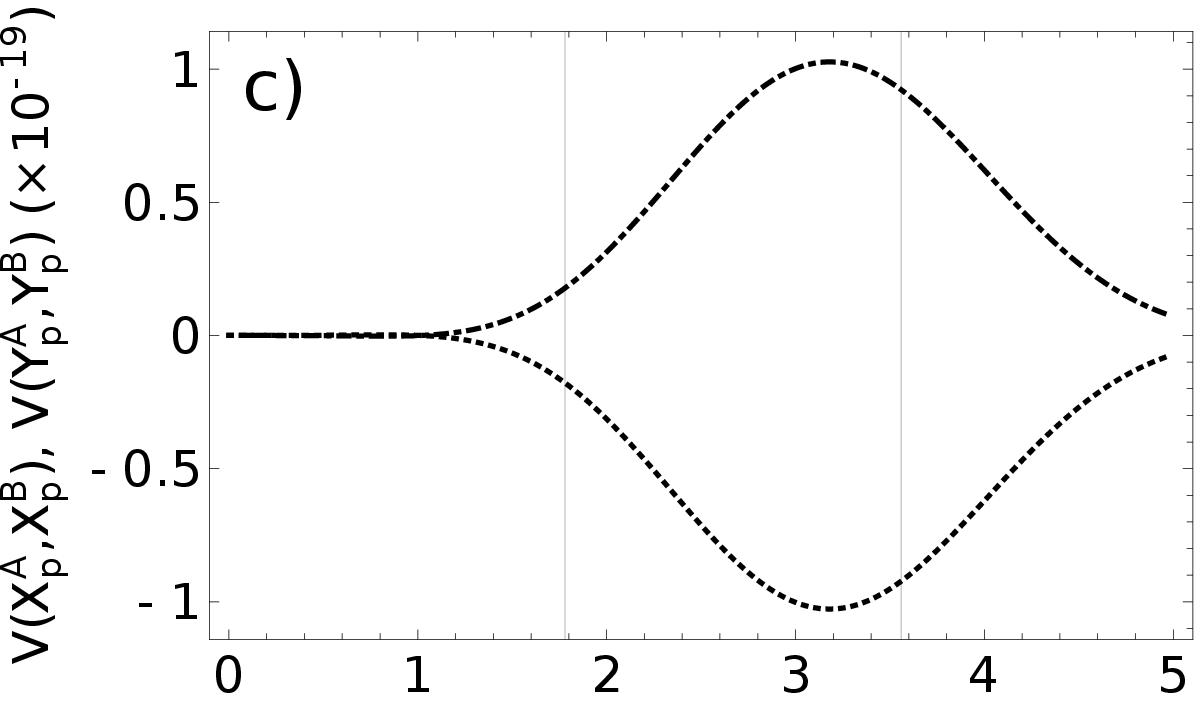}}
     \vspace {-0.2cm}
      \subfigure{\includegraphics[width=0.42\textwidth]{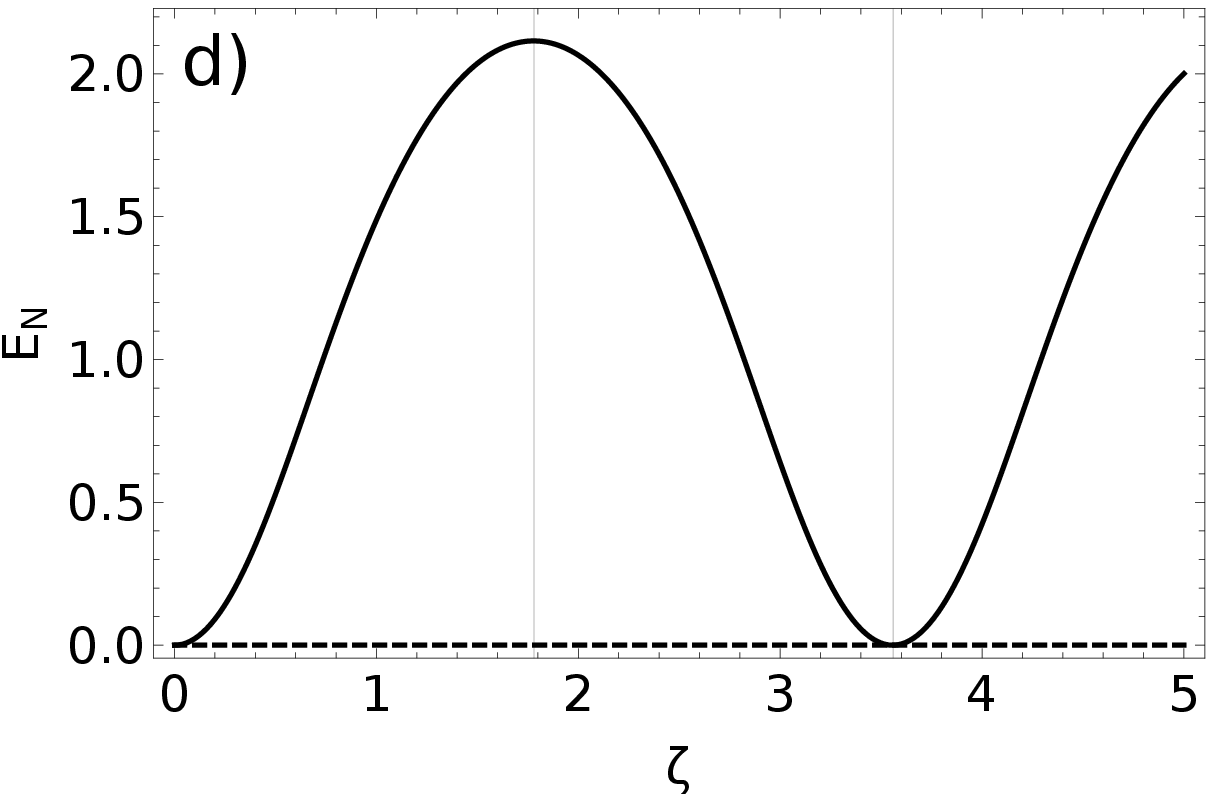}}
\vspace {0cm}\,
\hspace{0cm}\caption{\label{F2}\small{Negligible depleted-pump fields regime. From top to bottom: dimensionless  signal powers (solid) (a), the elements of the correlation matrix $V(X^{A}, X^{B})$ (dot), $V(Y^{A}, Y^{B})$ (dot-dash) for the signal (b) and pump (c) fields, respectively, and the logarithmic negativity $E_{\mathcal{N}}$ corresponding to the subsystem of signals (solid) and pumps (dash), respectively (d). $\kappa=1.13$ and $P_{s}/ P_{p}=10^{-20}$. $\zeta$ is the normalized propagation coordinate. The vertical lines show the planes where the phase difference of the classical waves $\Delta\theta$ ($\Delta \phi$) are multiples of $\pi$.}}
\end {figure}

\begin{figure}[p]
  \centering
    \subfigure{\includegraphics[width=0.425\textwidth]{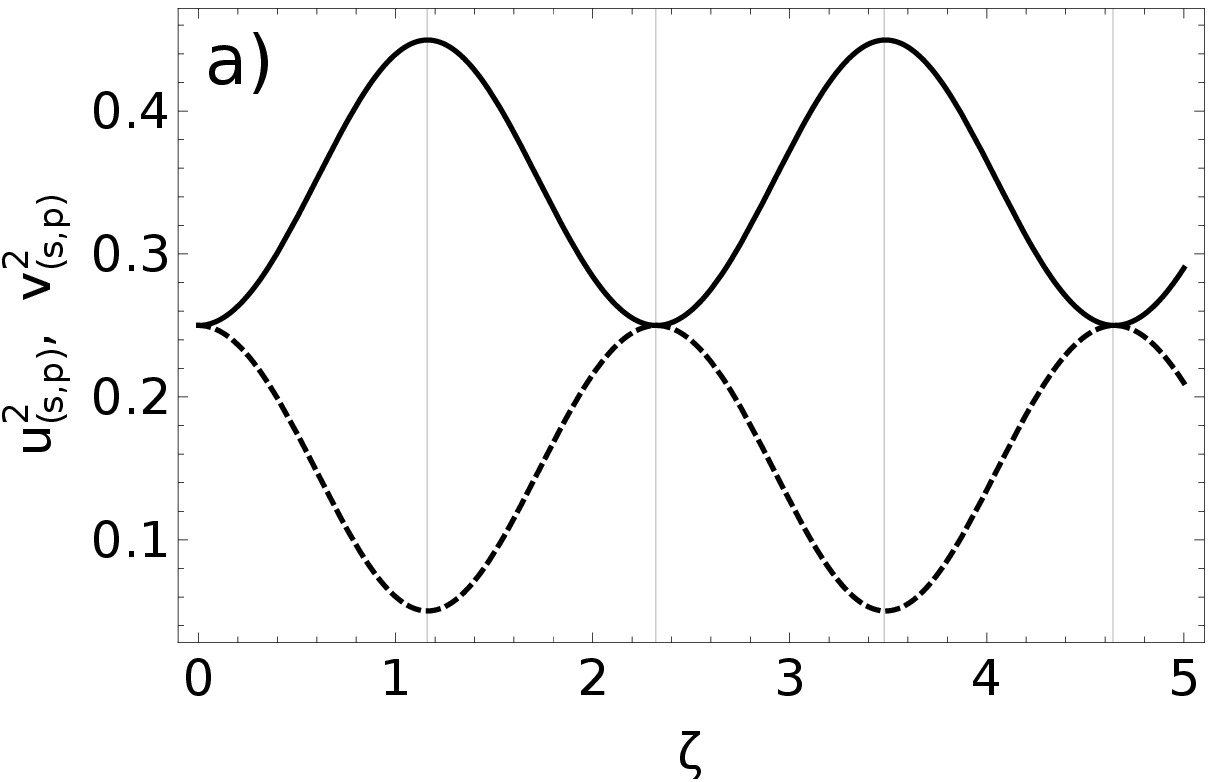}}
    \vspace {-0.2cm}
    \subfigure{\includegraphics[width=0.425\textwidth]{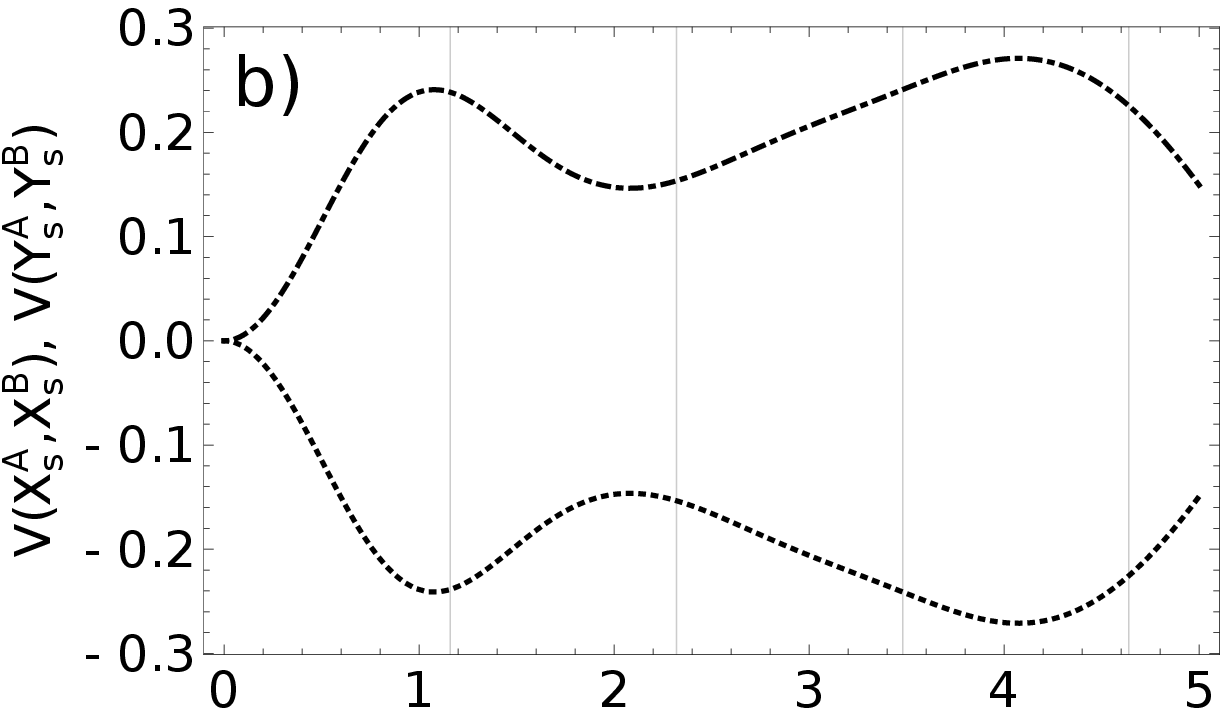}}
    \vspace {-0.2cm}
     \subfigure{\includegraphics[width=0.435\textwidth]{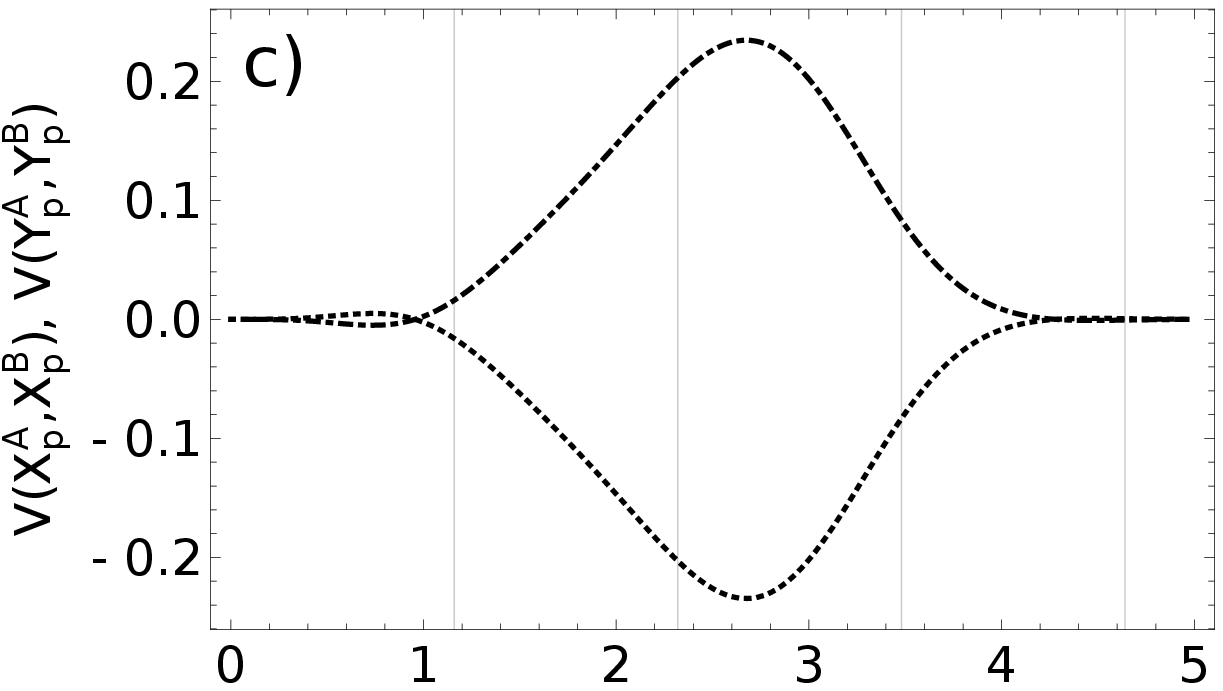}}
     \vspace {-0.2cm}
     \subfigure{\includegraphics[width=0.42\textwidth]{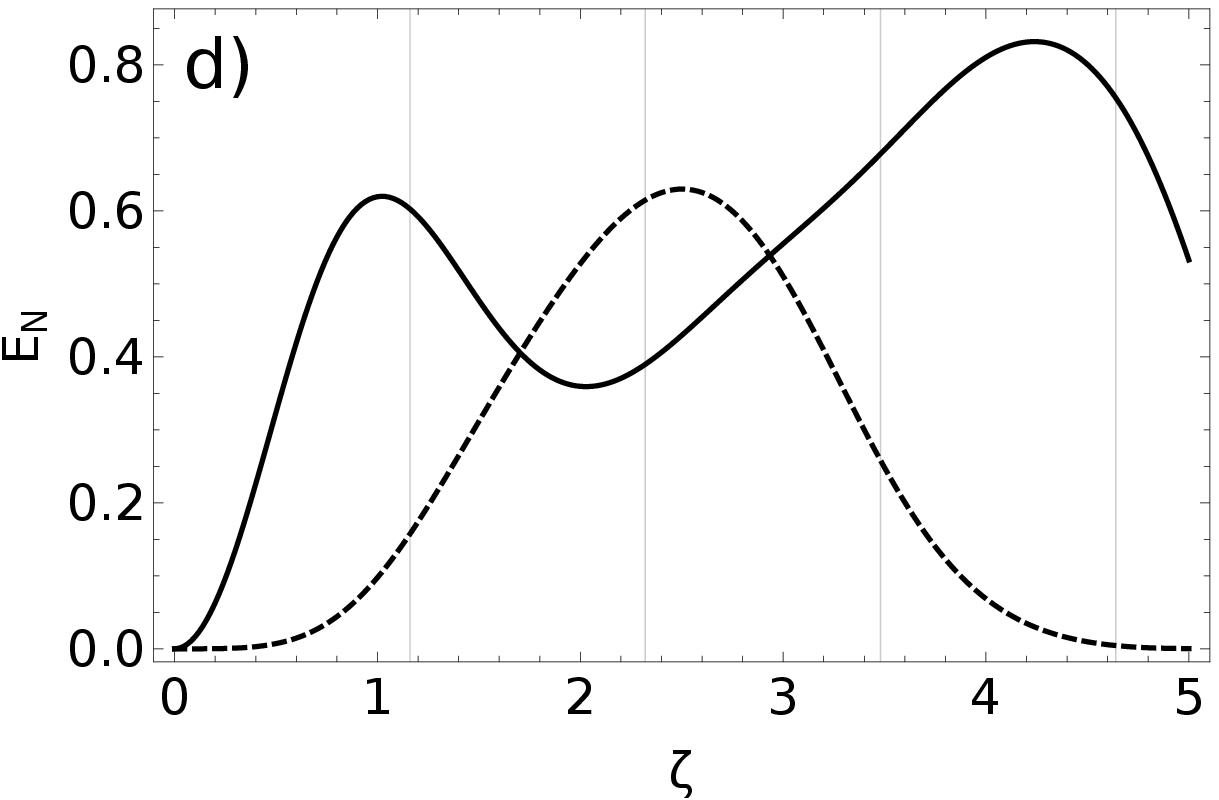}}
\vspace {0cm}\,
\hspace{0cm}\caption{\label{F3}\small{Highly depleted-pump fields regime. From top to bottom: dimensionless signal (solid) and pump (dash) powers (a), the elements of the correlation matrix $V(X^{A}, X^{B})$ (dot), $V(Y^{A}, Y^{B})$ (dot-dash) for the signal (b) and pump (c) fields, respectively, and the logarithmic negativity $E_{\mathcal{N}}$ corresponding to the subsystem of signals (solid) and pumps (dash), respectively (d). $\kappa=1.13$ and $P_{s}/ P_{p}=1$. $\zeta$ is the normalized propagation coordinate. The vertical lines show the planes where the phase difference of the classical waves $\Delta\theta$ ($\Delta \phi$) are multiples of $\pi$.}}
\end {figure}

In the negligible depleted-pump fields regime (Figure \ref{F2}), the classical dimensionless signal powers $u_{s}^{2}=v_{s}^{2}$ (Figure \ref{F2}a), the covariances of the signal waves (Figure \ref{F2}b) and the logarithmic negativity related to the signal modes (Figure \ref{F2}d), are driven by the difference of classical phases $\Delta\theta\,(\Delta \phi)$, as expected from the simplified calculation of section III, where vertical lines at each subfigure stand for multiples of $\pi$. In more details, coming back to  Equations (\ref{us}) - (\ref{tp}) it is observed that, due to the symmetry of the system, $\theta(\zeta)=\phi(\zeta)$ (numerically checked), so that the amplitudes depend only on the phase difference $\Delta \theta(\zeta)$, but not on the single phases. It is remarkable that the classical amplitudes are periodic oscillatory functions with $\Delta\theta$ independently of the depletion, as shown in Figures \ref{F2}a and \ref{F3}a.

 Besides, by closely inspecting Equations  (\ref{Quad1}) - (\ref{Quad4}), it can be easily seen that for $u_{s}(\zeta)\ll u_{p}(\zeta)$ (equivalent to getting rid of the pump fluctuations), $u_{p}(\zeta) \approx 1/\sqrt{2}$ and $\theta_{p}(\zeta) \approx 0$. Under this approximation, the solutions for the propagation of the signal quadratures lead to analytical expressions equivalent to Equations (\ref{US1}-\ref{US2}) of section III. Then, the elements of the covariance matrix under scrutiny for the signals read
\begin{equation}
V(X_{s}^{A},X_{s}^{B})=-V(Y_{s}^{A},Y_{s}^{B})\approx
-\frac{2^{3/2}\kappa \tilde{L}_{ab}^{2}}{\pi^2} \sin^{2} (\frac{\pi \zeta}{2 \tilde{L}_{ab}} ),
\end{equation}
with $\tilde{L}_{ab}=\pi/ (2 \sqrt{\kappa^{2} - (1/2)})$ the normalized beat length. This equation shows the strong anticorrelations appearing in the signals subsystem. For SPDC, $\zeta / \tilde{L}_{ab} \approx z / L_{ab}$, such that the covariances present the same periodicity as that corresponding to the number of photons given by Equation (\ref{Number}). 

Conversely, the covariance matrix elements related to the pump fields are very low, leading to negligible values of entanglement, and do not show the same periodicity (Figure \ref{F2}c). These values are zero in the approximation introduced above ($u_{s}(\zeta)\ll u_{p}(\zeta)$), as well as by using the theory of section III. In order to get the required accuracy on the covariance matrix elements ($10^{-19}$), the fluctuations of the pump fields have to be taken into account, allowing only a numerical solution of the Equations (\ref{Quad1}) - (\ref{Quad4}). Both the delay and different periodicity shown in Figure \ref{F2}c are due to the inclusion of the pump fluctuations ($\hat{X}_{p}, \hat{Y}_{p}$) in the calculation. 

Finally, notable values of logarithmic negativity for the signal as high as $2$ can be obtained in devices of less than $30$ mm \cite{Mallon2008}.

\subsubsection{High depletion}

Let us now analyze the highly depleted-pump fields regime (Figure \ref{F3}). The classical dimensionless powers oscillate periodically above ($u_{s}^{2}, v_{s}^{2}$) and below ($u_{p}^{2}, v_{p}^{2}$) the normalized initial value $1/4$  (Figure \ref{F3}a). Moreover, the signal covariances decrease as the pump covariances increase, reaching measurable values $V({\xi}_{j}, {\xi}_{k})>10^{-3}$ (Figures \ref{F3}b and \ref{F3}c) \cite{Buono2010, Dauria2009}. This rise brings as a consequence the entanglement of both the signal and, remarkably, the entanglement of the pump modes  which have not directly interacted (Figure \ref{F3}d). Actually, the seeding of the signal waves acts as an entanglement switch for the pump fields. It is important to outline that there are distances where values of entanglement as high as $1/2$ are found in both signal and pump subsystems. These values are on the order of those reported with schemes of optical cavities  \cite{Mallon2008}. A very interesting consequence of this effect is that the measurement of entanglement on one subsystem, signal or pump field, can be used as a non-perturbative measure of entanglement on the other. 

It should be noted that in this regime, the quantum features of the system are no longer fully driven by the classical phase difference $\Delta \theta$: a depletion-based phase mismatch drives the entanglement evolution (Figure \ref{F3}). In this case the relation of the cascade effect with the correlations gets more complex. Since both the pump and signal amplitudes and phases are on the same order, the auto-interaction of the signal quadratures given by the $u_{p} \sin(\theta_{p})$ factors in  Equations  (\ref{Quad1}) - (\ref{Quad2}) are no longer negligible. They are at the origin of the phase mismatch between classical and quantum propagation shown in Figure \ref{F3}. Furthermore, note that unlike the signals subsystem, where $V(X_{s}^{A},X_{s}^{B})$ and $V(Y_{s}^{A},Y_{s}^{B})$ are the only relevant elements, in the pumps subsystem other elements like  $V(X_{p}^{A},Y_{p}^{B})$ and $V(Y_{p}^{A},X_{p}^{B})$ (not shown) yield also non-negligible correlations which $E_{\mathcal{N}}$ also takes into account.

\begin{figure}[h]
  \centering
    \subfigure{\includegraphics[width=0.42\textwidth]{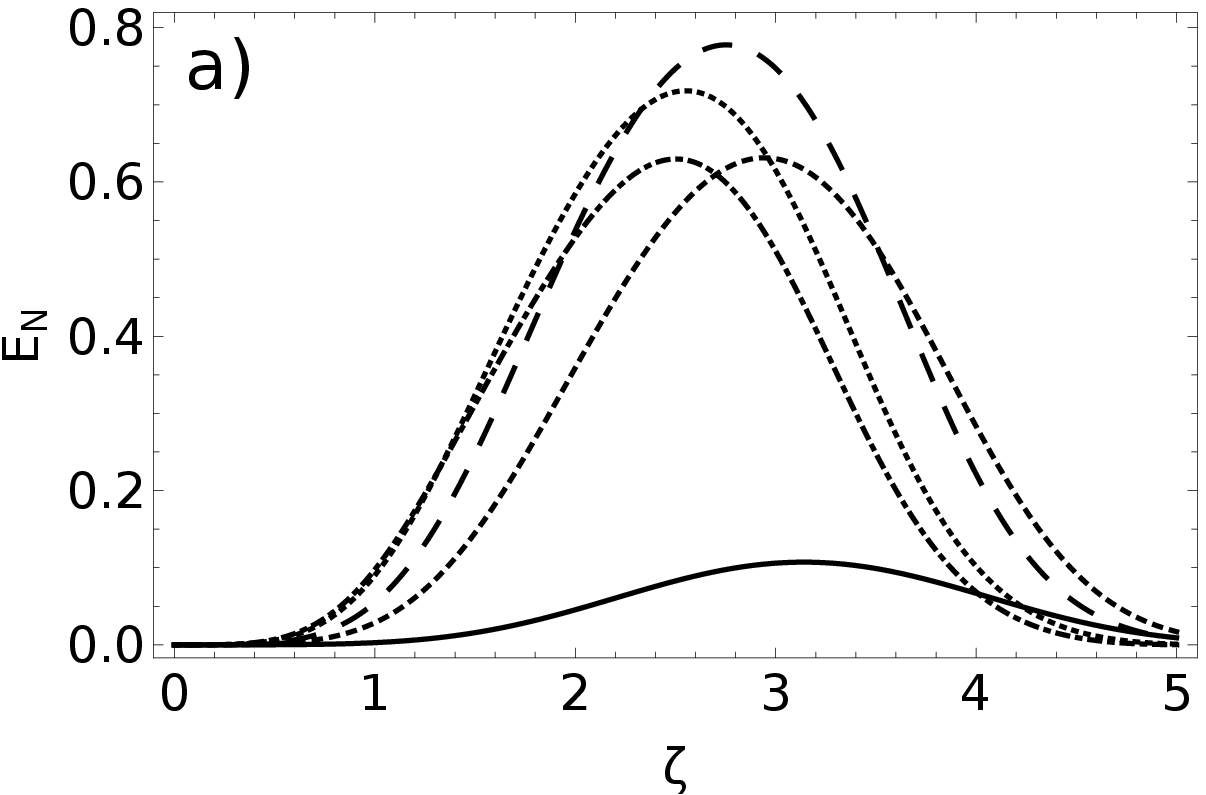}}
    \vspace {-0.2cm}
    \subfigure{\includegraphics[width=0.42\textwidth]{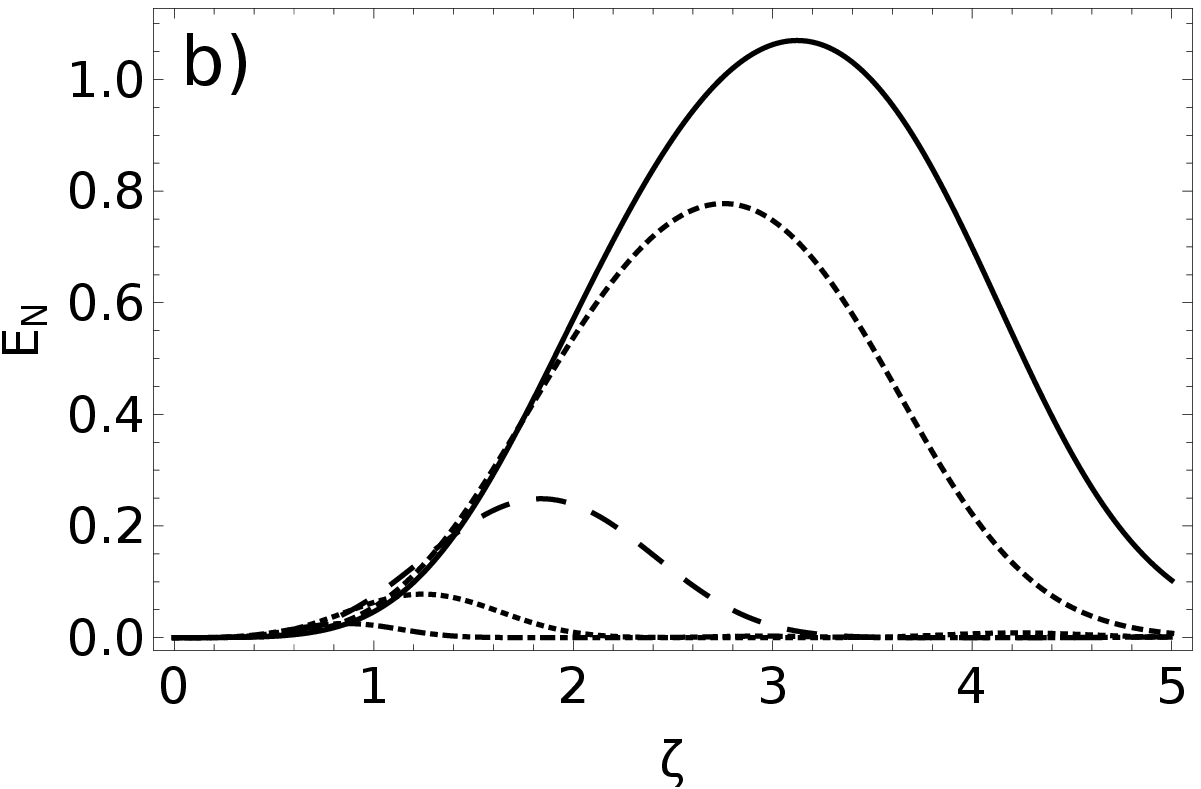}}
\vspace {0cm}\,
\hspace{0cm}\caption{\label{F4}\small{Pumps entanglement for different values of $\kappa$ and ratio $P_{s}/P_{p}$. Top (a): logarithmic negativity $E_{\mathcal{N}}$ for a set value of $\kappa=1.13$ and different ratios $P_{s}/P_{p}$: $1/10$ (solid),  $1/9$ (small-dash), $1/4$ (large-dash),  $2/3$ (dot) and  $1$ (dot-dash). Bottom (b): logarithmic negativity $E_{\mathcal{N}}$ for a set value of the ratio $P_{s}/P_{p}=1/4$ and different values of $\kappa$: $1.01$ (solid), $1.13$ (small-dash), $1.6$ (large-dash), $2.26$ (dot) and $3.2$ (dot-dash). $\zeta$ is the normalized propagation coordinate. Note that in the bottom subfigure, $\zeta$ is different for each plot. 
}}
\end {figure}

Let us now analyze more deeply the entanglement of the pump modes. Figure $\ref{F4}$a  shows the values of logarithmic negativity $E_{\mathcal{N}}$ versus propagation for different ratio of signal-pump input powers $P_{s}/P_{p}$ and a fixed value of the effective coupling $\kappa$. All the curves display a maximum close to $\zeta=3$. However, the global maximum is obtained for  $P_{s}/P_{p}=1/4$, which optimises the pump depletion. It is important to outline that the entanglement is obtained over a broad range of values of $\zeta$, in such a way that the length of the coupler is not a critical parameter. This is a fact of great practical importance when designing a sample. In Figure $\ref{F4}$b, we set $P_{s}/P_{p}=1/4$ and the different curves are obtained for different values of $\kappa$, i.e. the total power per waveguide. It is shown that the entanglement negativity follows the input power ($\kappa \rightarrow 1$), reaching its maximum for $\kappa=1.01$, which corresponds to the largest value of total input power considered here. In other words,  for a given nonlinearity $g$ and fixed total input power $P$, the lower the linear coupling $C$, the higher the entanglement strength. The reason is that the rate of down/up-conversion scales with the $\sqrt{2P} g$ factor  whereas the linear coupling controls the speed of cascading, such that if the coupling is low, more down-converted signal photons are available to be up-converted back to the pump modes and get entangled. Note that the normalized coordinate $\zeta$ is different for each plot, due to its dependence on the input power. In terms of physical lengths, the maxima of  entanglement are obtained for lengths $z=34$ mm with $\kappa=3.2$ (dot-dash) and $z=39$ mm with $\kappa=1.01$ (solid), both values being easily attainable in {lithium niobate}. 

\subsection{Quadripartite entanglement}

\begin{figure}[h]
\centering
\includegraphics[width=0.45\textwidth]{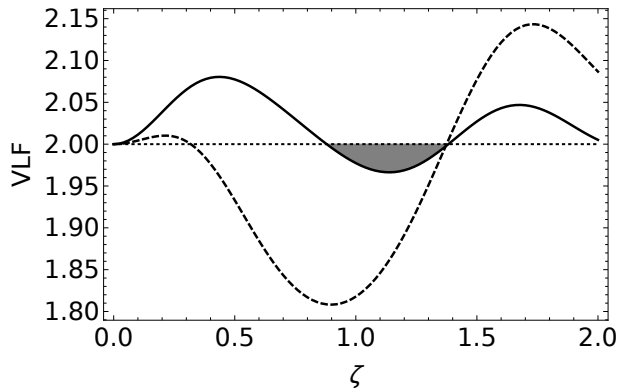}
\vspace {0cm}\,
\hspace{0cm}\caption{\label{F5}\small{Optimised van Loock - Furusawa inequalities. Simultaneous values under the threshold value imply CV quadripartite entanglement. Solid line: the first and third inequalities. Dash line:  the second inequality. Dot line: quadripartite entanglement threshold. In gray the area where the violation of the three inequalities is obtained. $\kappa=2.26$ and $P_{s}/P_{p}=1$. $\zeta$ is the normalized propagation coordinate.}}
\end {figure}

Eventually, we demonstrate the existence of quadripartite entanglement in the system under study. Figure \ref{F5} shows the van Loock and Furusawa inequalities given by Equations (\ref{VLF}) for an effective coupling $\kappa=2.26$ and $P_{s}/P_{p}=1$. The normalized propagation coordinate is such that $\zeta=1$ is equivalent to $z\approx 28$ mm. The parameters $r_{j}$ (j=1, \dots, 4) have been optimised to maximize the violation of Equations (\ref{VLF}). Due to the symmetry of the system, the violation of the first and third inequalities are equal (solid). Notably, there are lengths over which all the inequalities are violated, therefore showing two-colour quadripartite entanglement within the system (Figure \ref{F5}, gray area). Note that the degree of violation of the three inequalities is much lower than that obtained in cavities \cite{Migdley2010}. However, as highlighted above, our nonlinear directional coupler does not couple the pump fields through the evanescent coupling, so the mere appearance of this effect, even though weak, is outstanding. Furthermore, it is obtained for relatively low input powers and interaction lengths of about $30$ mm.

\subsection{Feasibility {and applications}}

We would like to outline that losses could be straightforwardly included in our analysis by following ref. \cite{Herec2003}, where it was shown that entanglement is maintained even under strong damping. Typical values of losses in PPLN waveguides are on the order of $0.1$ dB\,cm$^{-1}$ \cite{Alibart2016} which, under the assumptions of ref. \cite{Herec2003}, would lead to a reduction of approximately $6\%$ ($12\%$) in the $2$- ($4$-) waves interaction strength after $30$ mm of propagation within the device. These figures guarantee the suitable operation of the proposed device with the waveguide parameters $C$ and $g$ considered above. Further optimisation of these parameters for a requested operation mode, leading to either bipartite or quadripartite entanglement, can maximise the performance of the device.

Overall, all the results presented in this article could be obtained for reasonably low pump and seed powers by using already existing technology. Note that strong depletion in the traveling-wave regime usually requires high levels of pump power. Here it is not the case due to two main reasons. First, the depletion can be governed by the ratio of signal-pump input powers  $P_{s}/P_{p}$. This property was discussed at the classical level in reference \cite{Russel1993}. Second, the periodical funneling of the signal waves between the waveguides increases the depletion effect. 

{As closure, we outline that bright sources of bipartite entanglement can be of primary importance in sereval CV-based quantum protocols such as quantum communications \cite{Ralph1998, Ralph2009}, quantum key distribution \cite{Bencheikh2001, Silberhorn2002}, optomechanical entanglement \cite{Mazzola2011} and quantum imaging \cite{Boyer2008}, among others. Multipartite entanglement of bright beams opens up additional interesting avenues, such as multipartite EPR steering \cite{Armstrong2015}.}

\section {Conclusion}

We have studied the CV entanglement in a nonlinear $\chi^{(2)}$ directional coupler and have shown that two non-interacting bright pump fields become entangled during the propagation due to a nonlinear cascade effect. Likewise, we found that there are distances where both subsystems, pump and signal modes, show large values of entanglement, in such a way that the measurement of entanglement on one subsystem can be used as a measure of entanglement on the other, without destroying it. Moreover, we have shown that two-colour quadripartite entanglement is also present in the system under certain conditions. Finally, we have demonstrated that the device proposed here can be realized with current technology. Consequently, it stands as a good candidate for a source of multi-colour and/or multipartite entangled states for complex continuous-variable quantum information processing protocols.

\appendix*
\section{}

To clarify the origin of the covariance matrix elements shown in Figures $\ref{F2}$ and $\ref{F3}$, we include here the complete covariance matrix related to our system
\begin{widetext}
\begin{equation*}
\mathbf{V}=
\begin{pmatrix}
V(X_{s}^{A},X_{s}^{A}) & V(X_{s}^{A},Y_{s}^{A}) & V(X_{s}^{A},X_{p}^{A}) & V(X_{s}^{A},Y_{p}^{A}) & \mathbf{V(X_{s}^{A},X_{s}^{B})} & V(X_{s}^{A},Y_{s}^{B}) & V(X_{s}^{A},X_{p}^{B}) & V(X_{s}^{A},Y_{p}^{B}) \\
 & V(Y_{s}^{A},Y_{s}^{A}) & V(Y_{s}^{A},X_{p}^{A}) & V(Y_{s}^{A},Y_{p}^{A}) & V(Y_{s}^{A},X_{s}^{B}) & \mathbf{V(Y_{s}^{A},Y_{s}^{B})} & V(Y_{s}^{A},X_{p}^{B}) & V(Y_{s}^{A},Y_{p}^{B}) \\
 &  & V(X_{p}^{A},X_{p}^{A}) & V(X_{p}^{A},Y_{p}^{A}) & V(X_{p}^{A},X_{s}^{B}) & V(X_{p}^{A},Y_{s}^{B}) & \mathbf{V(X_{p}^{A},X_{p}^{B})} & V(X_{p}^{A},Y_{p}^{B})\\
 &  &  & V(Y_{p}^{A},Y_{p}^{A}) & V(Y_{p}^{A},X_{s}^{B}) & V(Y_{p}^{A},Y_{s}^{B}) & V(Y_{p}^{A},X_{p}^{B}) & \mathbf{V(Y_{p}^{A},Y_{p}^{B})} \\
&  &  &  & V(X_{s}^{B},X_{s}^{B}) & V(X_{s}^{B},Y_{s}^{B}) & V(X_{s}^{B},X_{p}^{B}) & V(X_{s}^{B},Y_{p}^{B}) \\
&  &  &  &  & V(Y_{s}^{B},Y_{s}^{B}) & V(Y_{s}^{B},X_{p}^{B}) & V(Y_{s}^{B},Y_{p}^{B}) \\
&  &  &  &  &  & V(X_{p}^{B},X_{p}^{B}) & V(X_{p}^{B},Y_{p}^{B}) \\
&  &  &  &  &  & & V(Y_{p}^{B},Y_{p}^{B}) 
\end{pmatrix},
\end{equation*}
with the relevant elements in bold. {Repetitive entries have been omitted since $\mathbf{V}=\mathbf{V}^{T}$}.
\end{widetext}


\section*{Acknowledgements}
D.B. would like to thank J. Li\~nares for stimulating discussions on the topic. This work was supported by the Agence Nationale de la Recherche through the INQCA project (grant agreement number PN-II-ID-JRP-RO-FR-2014-0013) and the Investissements d'Avenir program (Labex NanoSaclay, reference ANR-10-LABX-0035).

\end{document}